# Electric Levitation Using Epsilon-Near-Zero Metamaterials

Francisco J. Rodríguez-Fortuño[1,2], Ashkan Vakil[1] and Nader Engheta*[1]

[1] University of Pennsylvania. Department of Electrical and Systems Engineering. Philadelphia, Pennsylvania 19104, USA

[2] Nanophotonics Technology Center. Universitat Politècnica de València. 46022 Valencia, Spain

*Correspondence to: engheta@ee.upenn.edu

**Levitation of objects with action at a distance has always been intriguing to humans. Several ways to achieve this[1], such as aerodynamic, acoustic, or electromagnetic methods, including radiation pressure[2], stable potential wells[3], and quantum Casimir-Lifshitz forces[4], exist. A fascinating approach for levitation is that of magnets over superconductors based on the Meissner effect —the expulsion of the magnetic field by a superconductor[5]. With the advent of metamaterials— designed structures with electromagnetic properties that may not be found in nature—we ask whether a material may be conceived exhibiting similar field expulsion, but involving the electric field. We show how a special subcategory of metamaterials, called epsilon-near-zero materials[6–9], exhibits such electric classic analog to the Meissner effect, exerting a repulsion on nearby sources. Repulsive forces using anisotropic and chiral metamaterials have been investigated[10–16], but our proposal uses a different mechanism based on field expulsion, and is very robust to both losses and material dispersion.**

Figure 1 illustrates the basis of our proposal for electric levitation in vicinity of an ENZ metamaterial. Figure 1(a) displays a magnet over a superconducting substrate, demonstrating, in a simple form, the well-known physics of the Meissner effect. The superconductor expels the magnetic flux density **B** as it does not allow the flux to penetrate inside. As this is a quantum phenomenon, the superconductor does not simply behave as a perfect conductor, implying only the condition $\partial \mathbf{B}/\partial t = 0$, but acts analogous to a perfect diamagnetic material in which **B** = 0. Although neither of the hypotheses of magnetic permeability μ being zero and conductivity σ → ∞ offers a complete description of the transition into the superconducting state, the phenomenon is occasionally associated[17] with μ = 0—one would expect that a material with μ = 0 expels the magnetic fields away from its surface (since normal component of the magnetic flux density **B** should be zero outside the material near its surface). This will exert a magnetostatic gradient force on the magnet and levitate the magnet to a stable point. Inspired by this notion, we envision an analogue classical scenario, shown in Fig. 1(b). Can we have an electric classical dual of Meissner effect involving metamaterials? In other words, would an epsilon near zero material (ENZ) exhibit a similar behaviour, albeit classical, for the electric fields? In an ENZ substrate, the displacement current **D** generated by a polarized particle will be expelled in an analogous (but classic) fashion, and thus may result in levitation of the particle. The apparent simplicity of this analogy may be misleading: one could argue that the same analog to Meissner effect could be achieved with the use of a classic conductor, inside which the electric field is zero. However, in such case free charges would accumulate on the surface of a conductor, creating an outside electric field normal to the surface of the conductor which results in a different field profile than that of Fig. 1(b), resulting in attraction rather than levitation. In contrast, an ENZ material does not accumulate free charges inside its bulk or at its boundary. The normal component of the electric displacement **D** is therefore conserved at the interface, as in for any finite-conductivity dielectric. This means that analogous to a superconductor, in which the magnetic flux **B** is zero inside and therefore—due to continuity of the normal components—it must be parallel to the surface of the superconductor outside, the electric displacement **D** at the boundary of an ENZ material must also necessarily be parallel to the surface outside (and of course zero inside since the permittivity is zero), creating a complete analogy (again albeit classical) between both field distributions. Since ENZ materials cannot exist at zero frequencies, the proposed mechanism works for a source oscillating at an angular frequency ω, as Fig. 1(b) depicts, but the static analogy is still valid for the near fields and for forces at electrically small distances. For values of permittivity close to, but not exactly zero, the effect still exists, and is an electric analog to conventional diamagnetic repulsion.

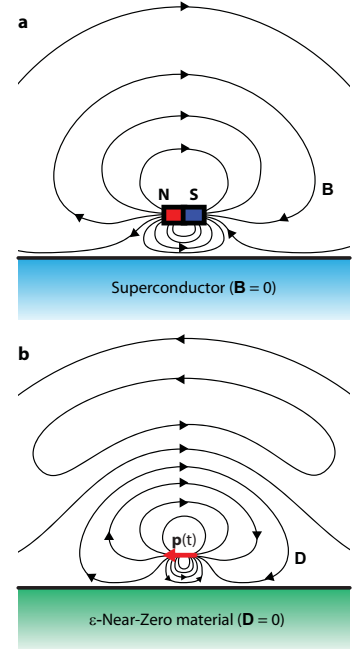

**Fig. 1.** *ENZ analog to the Meissner effect. Analogy between (a) the Meissner effect, where the magnetic field created by a permanent magnet cannot penetrate into a superconductor substrate, leading to a levitation of the magnet, and (b) the levitation of an electric radiating dipole source over an ε-Near-Zero (ENZ) substrate, into which the electric displacement field cannot penetrate.*

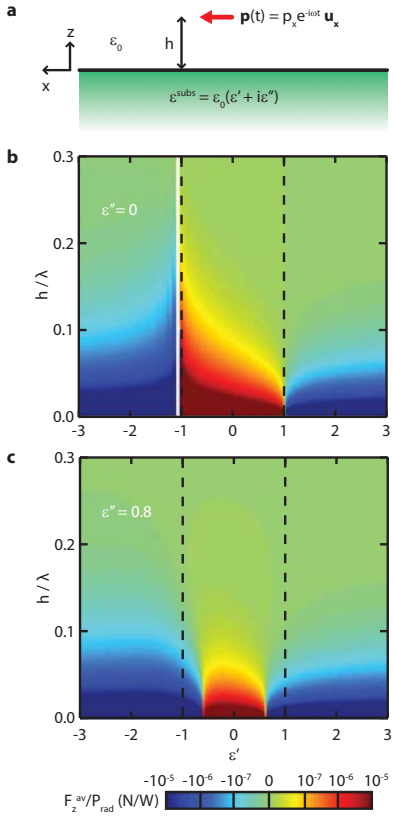

**Fig 2**. *Repulsive force dependence on height and substrate permittivity* (a) Geometry of the problem. (b,c) Plot of the time-averaged force acting on a horizontal electric point dipole, per unit radiated power of the dipole, as a function of the height over the substrate (vertical axis) and the real part of the permittivity of the substrate (horizontal axis). The imaginary part of the permittivity of the substrate is set to zero (lossless) in (b) and set to a very high value $\varepsilon'' = 0.8$ in (c). In both cases a repulsive force is observed around $\varepsilon' = 0$.

To understand the physics behind our proposal, we consider an idealized and simple scenario as depicted in Fig. 2(a). We begin our study by considering a point electric dipole oscillating monochromatically with an angular frequency $\omega$ above a homogeneous semi-infinite substrate. We first calculate the total electric field created by the dipole and "reflected" by the substrate, applying the well-known Sommerfeld integral[18–21], which yields a rigorous solution to the Maxwell equations (details given in supplementary information). The resulting electric field distribution is exactly the expected one shown in Fig. 1(b) (see movie S1 in the supplementary information). This is strong evidence that a repulsive force exists, but we would like to calculate it analytically. We computed the time-averaged force due to the reflected electric field upon the dipole as a function of the height $h$ of the dipole above the substrate, and as a function of the complex permittivity of the substrate $\varepsilon_{subs} = \varepsilon_0(\varepsilon'+i\varepsilon'')$. Contrary to the static nature of the Meissner effect, our scenario deals with time-varying fields, so the complete expression for the force **F** acting on a stationary time-varying dipole **p** is given by $\mathbf{F} = (\mathbf{p}\cdot\nabla)\mathbf{E} + (\partial\mathbf{p}/\partial t)\times\mathbf{B}$, including a term due to the inhomogeneous electric field, and a term due to the Lorentz force acting on the charges moving in a magnetic field[19]. The resulting net force, after time-averaging, can be expressed as an indefinite integral, which can be evaluated numerically (the details of this calculation are given in the supplementary information). The resulting time-averaged force $F_z^{av}$ is shown to be directed vertically and varies linearly with the radiated power of the dipole $P_{rad}$ in unbounded free space. The resulting time-averaged net force per unit radiated power ($F_z^{av}/P_{rad}$) turns out to be independent of the oscillating frequency of the dipole for a fixed $(h/\lambda)$, where $\lambda$ is the wavelength of the radiation. In Fig. 2, (b) and (c), this force per unit radiated power acting on a horizontal electric dipole (HED) is shown as a function of $(h/\lambda)$ and $\varepsilon_{subs}$. The analogous results for a vertical dipole, provided in the supplementary information, show a similar behavior. Any reorientation of the dipole, expected due to the existence of a torque, will be a superposition of the horizontal and vertical scenarios, and will also experience a levitating force.

Let us consider the lossless case first (Fig. 2(b)). The color in the figure represents the magnitude of the vertical time-averaged force acting on the dipole per unit radiated power; negative (blue) values represent attraction toward the substrate, while positive (red) values represent repulsion away from the substrate. A zero value (green color) implies that there is no time-averaged force acting on the dipole. As can be seen, when the substrate is a conventional dielectric ($\varepsilon' > 1$), there is an attractive force acting on the dipole: this is well-known and it is interpretable as the force on the dipole due to its image. This force, proportional to the radiated power of the dipole, increases as the dipole approaches the surface of the substrate. When the permittivity of the substrate is equal to that of air ($\varepsilon' = 1$), the dipole is radiating in free space and there is no force acting on the dipole. The interesting effect is observed when we further reduce the permittivity, in the region $-1 < \varepsilon' < 1$, where a repulsive force exists on the dipole. In particular, when the substrate permittivity is near zero ($\varepsilon' \approx 0$), the electric displacement field radiated by the dipole cannot penetrate into the substrate, leading to the field distribution shown in Fig. 1(b). The repulsive force in this case is the classic analog to the repulsive force acting on a magnet over a superconductor, exhibiting an analogous ENZ-based Meissner-like effect. This can be understood as a force pushing the system away from the high-energy configuration that exists when the dipole is very close to the substrate, associated with very high values of the electric field, "squashed" between dipole and substrate. It is worth noting that a resonant force exists when $\varepsilon'$ is close to $-1$. Such resonant force is result of surface plasmon resonances at the boundary between air and substrate[16]. This resembles the high forces that arise using magnetostatic surface resonances for levitation of metamaterials[22]. Ideally, this resonance can result in an arbitrarily large force acting on the dipole, but as is typical with resonant phenomena, it is limited in practice by the losses.

However, the repulsion mechanism at the ENZ condition ($\varepsilon' = 0$) does not rely on the existence of a resonant mode but on a property of the substrate at a particular frequency, leaving us to expect that such repulsion may be more robust to losses. That is indeed the case, as shown in Fig. 2(c), which shows the force when the substrate has very high losses ($\varepsilon'' = 0.8$, high compared to practical ENZ materials, e.g., SiC is a natural ENZ material[23] at around 29 THz where it has $\varepsilon'' \approx 0.1$). We have deliberately chosen an exaggerated imaginary value of $\varepsilon'' = 0.8$ to illustrate our point. The repulsive force around $\varepsilon' = -1$ based on the surface plasmon resonance has disappeared completely, but the repulsion in the region around $\varepsilon' = 0$ is still strong despite the high imaginary part, showing that our approach is robust to loss.

The results discussed so far are numerical evaluations of the exact integral for the time-averaged force on the dipole. Agreement of the force calculated in this way with the formal calculation of the force using Maxwell's stress tensor integrated in a surface surrounding the dipole will be shown later, providing two independent verifications of the validity of our results. However, in both cases, these integrals may not be helpful in providing insights into the physics behind the behaviors portrayed above. Moreover they are not easy to calculate. Thus it is convenient to have at our disposal a simple analytical expression. To this end, by applying the "quasistatic" (i.e, near-field) approximation to the integral for the force and by retaining the dominant term at low heights, where the force is most strong, we arrive at a simple expression for the time averaged force (see supplementary information for more details)

$$F_z^{ave} \approx K \cdot P_{rad} \cdot \left(\frac{h}{\lambda}\right)^{-4} \qquad (1)$$

where $K = (9/512\pi^4 c) \cdot p \cdot \Re[(\varepsilon_0 - \varepsilon_{subs})/(\varepsilon_0 + \varepsilon_{subs})]$, in which $p = 1$ and $2$ for a horizontal and vertical dipole, respectively. For a horizontal dipole over a ENZ substrate, $K$ is a constant given approximately by $+6\times10^{-13}$ $s \cdot m^{-1}$. The quasi-static approximation employed turns out to be equivalent to the image theory for static charges over a substrate[24]. Conveniently, we can infer all of the behavior already discussed above from this simple expression—e.g., the fact that in the lossless case the force is repulsive for $-1 < \varepsilon' < 1$, the divergence of the force at $\varepsilon' = -1$, which is degraded by the losses, and the fact that the repulsive force is robust to losses around $\varepsilon' = 0$. As can be seen, the repulsive force will only be strong for heights much smaller than the wavelength, so we are dealing with a near-field effect, and the repulsive force decays strongly with the height. Furthermore, since the expression is frequency independent (aside from the $(h/\lambda)^{-4}$ term), we can consider the locus of all points in the ($\varepsilon'$, $\varepsilon''$) space that result in a repulsive force, from which we can obtain the associated frequency bandwidth of repulsion for any dispersive ENZ substrate as $\varepsilon(\omega)$ moves through this region. For the Lorentz model of SiC given in Ref [23], a relatively wide repulsive force fractional bandwidth of 6% around the 29 THz ENZ frequency is obtained. This wide bandwidth is thanks to the broad range of values around ENZ that result in a repulsive force, which resembles conventional repulsion of diamagnetic materials, with permeability $\mu$ lower than 1, inside an inhomogeneous magnetic field. This resemblance may suggest that the electric levitation introduced here might be deemed as the *classic* imitation of Meissner effect, but at non-zero frequencies.

Figure 3 shows the time-averaged force on the electric dipole normalized to the radiated power in unbounded free space, as a function of $(h/\lambda)$ when the substrate is ENZ ($\varepsilon' = 0$). It shows both the numerical result of the exact integral expression for the force and the approximation given by Eq. (1), which yields almost identical values for $(h/\lambda) < 0.2$. The effect of a realistic value for losses in the substrate ($\varepsilon'' = 0.1$ which corresponds to the losses of SiC at its ENZ frequency of 29 THz) is also included to show that the force is very robust to losses. Fig. 3 also provides numerical confirmation of the analytical results by plotting the force obtained from numerical simulations using the commercially available software package[25], CST Microwave Studio™. To do so we first calculated the fields around the dipole over a realistic lossy ENZ substrate, and then post process them to compute Maxwell's stress tensor. Finally we integrated the tensor across certain surfaces to obtain the force acting on the dipole (details of this procedure are described in the supplementary information).

So far we have only discussed the repulsive force acting on the electric dipole, but in order to achieve levitation, this force should be greater than the electric dipole's weight. The previous results are independent of the weight of the dipole and valid at any frequency. In a practical levitation application using ENZ, however, the nature and weight of the source will depend greatly on the frequency at which we want to work. Although we have discussed an ideal dipole, from the analogy with the Meissner effect we can foresee that any source, close enough to the ENZ surface, will experience a levitating force due to field expulsion, so the validity of the dipole approximation is not a critical concern for an experimental realization. Anyhow, the ideal dipole is useful to provide us with an estimate of the required conditions, so we can use the simple expression of the force given by (1) to explore different scenarios as discussed below.

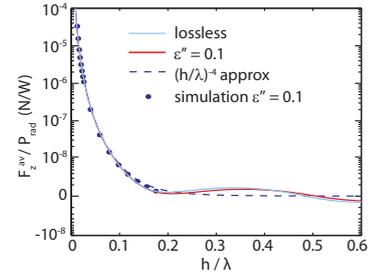

**Fig. 3** *Repulsive force dependence on height*. *Plot of the normalized time-averaged force acting on a horizontal electric point dipole over an ENZ substrate as a function of the normalized height. Both lossless and lossy cases are plotted. The approximation given by Eq. (1) in the text is also plotted as a dashed line, showing very good agreement at low normalized heights. Numerically simulated values of the force at different heights using Maxwell's stress tensor are also shown. The simulation points are only calculated for $(h/\lambda) < 0.2$, since for greater heights the simulation region became exceedingly large, and the numerical errors in the integration became comparable to the exact value of the vanishingly small force at greater heights.*

At microwave frequencies, electrically small sources and distances are relatively easier to work with; therefore we can exploit the relatively big repulsion forces at very low heights by considering electrically small antennas placed very close to an ENZ substrate. Let us assume a dipole cylindrical antenna of length $\lambda/4$ and diameter $\lambda/100$ made of copper: this assumption is only used to calculate the weight of the antenna and the dipole moment as a function of the input current supplied. Substituting the values into (1) we obtain that to have repulsive force greater than the weight of the antenna, so that the antenna can be levitated, the current supplied to the antenna must be $I$ (A) > $(5.56×10^4$ A·s$^{-3/2})[f$ (GHz)$]^{-3/2}$ $(h/\lambda)^2$ (details provided in the supplementary information). For example at 1 GHz, a current of 6 A is sufficient to levitate the described antenna a distance $\lambda/100$ (i.e., 3 mm) from the ENZ substrate. Notice that this height is electrically very small, due to the fact that the weight to levitate is relatively big. The calculations presented here are only a rough estimate of the achievable levitating phenomena. So an experimental setup should use an electrically small antenna and as well should take into account the effect of the ENZ substrate in the input impedance of the antenna. Considering the possibility of an electric breakdown or discharges resulting from molecular ionization in intense electric fields, one may probably require high vacuum to levitate heavy weights, a limitation not present in the more common magnetic levitation techniques.

At optical frequencies, however, we can exploit the fact that particles are very light, so we are not only restricted to very electrically small heights and strong fields. For example, consider a fluorescent nanosphere of radius $R$ with a mass density of $\rho$, radiating at the ENZ frequency of the substrate with a power $P_{rad} = \eta P_{inc}$ proportional to the incident power from a laser pump at another wavelength, where $\eta$ is the efficiency and $P_{inc} = P_{laser} (\pi R^2)/A_{laser}$ is the power incident on the nanosphere from a laser with total power $P_{laser}$, illuminating a spot of area $A_{laser}$. Under those approximate assumptions and momentarily ignoring the optical force due to the laser beam, to levitate the particle (i.e. repulsion from the ENZ substrate caused by its fluorescent radiation greater than its weight) the required power must follow this inequality:

$$P_{laser} > A_{laser} \cdot (4/3) \cdot (g/c) \cdot (1/\eta) \cdot \rho \cdot R \cdot (h/\lambda)^4,$$

where $g$ is the acceleration rate of gravity. Substituting reasonable values for these parameters ($\rho = 3$ g/cm$^3$, larger than that of silicon or aluminum, $A_{laser} = \pi$ $(1mm)^2$, corresponding to a spot size of radius 1 mm, and $\eta = 1\%$) we arrive at the inequality

$$P_{laser}[W] > 2 \times 10^4 (W/nm) \cdot R(nm) \cdot (h/\lambda)^4$$

meaning that we could levitate a nanoparticle of radius 50 nm, at a height $(h/\lambda) = 0.1$, using a laser with $P_{laser} > 100$ W, directed at a spot of radius $1mm$ containing one or many of the fluorescent particles, each of them receiving a power $P_{inc} = 250$ nW, and emitting 1% of it. In a practical experiment, one should also account for the optical forces caused by the pump laser[26]. These forces are dependent on the specific frequencies and the material used as substrate. A self-consistent calculation of a polarizable particle illuminated by a plane wave over an ENZ substrate is included in the Supplementary Materials.

Finally it is worth mentioning that the nonresonant levitating force presented in this work is not localized in contrast with gradient forces. As such similar to some previously proposed levitation scenarios[27], the applications of our proposal are different from those of gradient forces used in optical tweezers, for which the force is localized to the focus of the tweezers."

In conclusion, we have proposed a method for levitating electromagnetic sources over an ENZ substrate. The mechanism is very robust to both realistic losses and to frequency detuning from the ENZ frequency (and thus robust to material dispersion), rare traits in metamaterial applications. Unlike the Meissner effect, which uses magnetostatic fields, this classical effect is based on the property of ENZ media regarding the electric displacement field, and it can operate at any oscillating frequency. ENZ behavior can be found in natural materials at certain frequencies or can be designed for specific frequencies using the notion of metamaterials. The effect may have useful potential applications, and we have shown realistic requirements for levitation showing that future experiments are within reach. Further studies can be pursued on this analogy to the Meissner effect, inspired by the concepts from superconductor levitation, such as the imitation of flux-pinning effects which take place in a Type-II superconductor[28], by introducing air channels on ENZ media.

**Acknowledgements:**

This work is supported in part by the US Office of Naval Research (ONR) Multidisciplinary University Research Initiative (MURI) grant number N00014-10-1-0942. F.J. Rodríguez-Fortuño acknowledges financial support from grant FPI of GV and the Spanish MICINN under contracts CONSOLIDER EMET CSD2008-00066 and TEC2011-28664-C02-02.

## Supplementary Materials:

*Calculation of the electric field generated by an electric dipole radiating over a substrate*

For the calculation of the electric field created by a point electric dipole source radiating over a substrate, we used Sommerfeld's approach[18]. To obtain the electric field, we use the Hertz potential vector as following

$$\mathbf{E} = \nabla(\nabla \cdot \mathbf{\Pi}) + \omega^2 \mu_1 \varepsilon_1 \mathbf{\Pi}, \tag{S1}$$

where $\varepsilon_1$ and $\mu_1$ are the permittivity and permeability of the medium filling the upper space above the substrate (notice that in the main text we take such medium to be vacuum, i.e. $\varepsilon_1 = \varepsilon_0$ and $\mu_1 = \mu_0$, but throughout the supplementary information we consider a more general scenario). In the space above the substrate, the Hertz vector potential can be divided into primary (radiated by the dipole) and secondary (reflected by the substrate) components

$$\mathbf{\Pi} = \mathbf{\Pi}_p + \mathbf{\Pi}_s, \tag{S2}$$

The primary term corresponding to a dipole radiating in free space is given by Sommerfeld *(12)* as

$$\mathbf{\Pi}_p(\mathbf{r}) = (p_x \hat{\mathbf{x}} + p_z \hat{\mathbf{z}}) \left[ \exp(ik|\mathbf{r} - \mathbf{r}'|) / (4\pi|\mathbf{r} - \mathbf{r}'|) \right]. \tag{S3}$$

Where $p_x$ and $p_z$ are the dipole moment components in the x and z directions. Note that we have considered zero dipole moment in the y direction, without loss of generality, since we can always choose our XZ coordinate plane to contain the polarization vector of the dipole. In our calculations we are assuming $e^{-i\omega t}$ time-harmonic variations.

The secondary term can be obtained as a superposition of cylindrical functions. By applying the boundary conditions for the Hertz potential at the interface between the substrate and the upper space, one can obtain the coefficients of such superposition, yielding the following integral for the secondary fields *(12)*:

$$\mathbf{\Pi}_s(\rho, \varphi, z) = \frac{-1}{4\pi\varepsilon_1} \int_0^\infty \frac{k_\rho}{ik_{z1}} \left( p_x r^s(k_\rho) \hat{\mathbf{x}} + p_z r^p(k_\rho) \hat{\mathbf{z}} \right) J_0(k_\rho \rho) e^{ik_{z1}(z+h)} dk_\rho + \\ \hat{\mathbf{z}} p_x \cos\varphi \int_0^\infty k_\rho^2 a(k_\rho) J_1(k_\rho \rho) e^{ik_{z1}(z+h)} dk_\rho, \tag{S4}$$

where

$$r^s(k_\rho) = \frac{\mu_2 k_{z1} - \mu_1 k_{z2}}{\mu_2 k_{z1} + \mu_1 k_{z2}}, \tag{S5a}$$

$$r^p(k_\rho) = \frac{\varepsilon_2 k_{z1} - \varepsilon_1 k_{z2}}{\varepsilon_2 k_{z1} + \varepsilon_1 k_{z2}}, \tag{S5b}$$

$$a(k_\rho) = -\frac{2}{k_1^2} \frac{(k_{z1}+k_{z2})(k_{z1}-k_{z2})}{(k_{z1}(\mu_2/\mu_1)+k_{z2})(k_{z1}(\varepsilon_2/\varepsilon_1)+k_{z2})},\qquad \text{(S5c)}$$

in which $J_n(x)$'s are the Bessel functions of the first kind, $r^s$ and $r^p$ are the Fresnel reflection coefficients for s-polarized and p-polarized waves, $k_{zi} = \sqrt{k_i^2 - k_\rho^2}$ (complex in general) is the wave number in the z direction, $k_\rho$ is the wave number in the radial direction (the variable over which the integration is performed), $\varepsilon_2$ and $\mu_2$ are the permittivity and permeability of the substrate, and $k_i = \omega\sqrt{\varepsilon_i \mu_i}$. We numerically calculate this integral by integration along the real axis, using standard numerical algorithms, with special care near the edges of the integrand singularities (such as $k_\rho = k_0$). We did not find it necessary to resort to integration techniques in the complex plane[21].

We numerically computed the field for a vertical and horizontal dipole over an ENZ substrate. This process involves using Eqs. (S1) to (S4), yielding a number of integration procedures. The resulting animated electric field for a dipole at a height $h = 0.1\lambda_1$ is provided in movies S1 and S2 and was used as a guide to create the field plot in Fig. 1(b)

*Calculation of the time-averaged force acting on the electric dipole*

To calculate the complete time-averaged force acting on the dipole, we compute the force caused by the secondary (reflected fields) on the time-varying dipole **p**[19]

$$\mathbf{F}(t) = (\mathbf{p}\cdot\nabla)\mathbf{E} + (\partial\mathbf{p}/\partial t)\times\mathbf{B} + (\partial\mathbf{r}/\partial t)\times(\mathbf{p}\cdot\nabla)\mathbf{B} \qquad \text{(S6a)}$$

where **E** and **B** denote the secondary (reflected) electric and magnetic fields (i.e. the total fields minus the radiated fields of the dipole) at the location **r** of the dipole. The first term denotes the force originating from the inhomogeneous electric field, the second term is the Lorentz force caused by the magnetic field on the moving charges, and the third term, related to the movement of the dipole in an inhomogeneous magnetic field, is zero for a stationary dipole (and is negligible for dipoles moving at speeds small compared to the speed of light). Dropping the third term, the remaining two terms after time-averaging and using $\partial\mathbf{B}/\partial t = -\nabla\times\mathbf{E}$, can be rearranged into a single term [19] given by:

$$\langle\mathbf{F}(t)\rangle = \sum_{i=x,y,z} \frac{1}{2}\Re\{p_i^* \nabla E_i\},\qquad \text{(S6b)}$$

Simple symmetry considerations imply that the total force acting on the dipole must be directed along the z-axis (a vertical force), so we consider only the z component

$$\langle F_z(t)\rangle = \frac{1}{2}\Re\left\{p_x^*\frac{\partial E_x}{\partial z} + p_y^*\frac{\partial E_y}{\partial z} + p_z^*\frac{\partial E_z}{\partial z}\right\}. \qquad \text{(S7)}$$

In order to compute this force, we only require knowledge of the secondary electric field on the z-axis. The Sommerfeld integral for the Hertz potential of the secondary field can be solved in the z-axis, and reduces, after mathematical manipulation, into a much simpler expression, from which we can obtain the electric field in the z-axis as an integral expression [Eq. S8]. Alternatively, the same expression can be easily derived using the plane wave decomposition described in[19] and solved in the z-axis only, which is a more intuitive approach. Both methods yield the same result

$$\mathbf{E}_{\text{sec}}(\mathbf{r}=0\hat{\mathbf{x}}+0\hat{\mathbf{y}}+z\hat{\mathbf{z}}) = \frac{-1}{8\pi\varepsilon_1}\int_0^\infty \frac{k_\rho}{ik_{z1}}\begin{bmatrix} k_1^2 r^s - k_{z1}^2 r^p & 0 & 0 \\ 0 & k_1^2 r^s - k_{z1}^2 r^p & 0 \\ 0 & 0 & 2k_\rho^2 r^p \end{bmatrix}\begin{bmatrix} p_x \\ p_y \\ p_z \end{bmatrix} e^{ik_{z1}(z+h)} dk_\rho. \quad (S8)$$

The derivative with respect to $z$ can be easily obtained as:

$$\frac{\partial \mathbf{E}_{\text{sec}}(\mathbf{r}=0\hat{\mathbf{x}}+0\hat{\mathbf{y}}+z\hat{\mathbf{z}})}{\partial z} = \frac{-1}{8\pi\varepsilon_1}\int_0^\infty k_\rho \begin{bmatrix} k_1^2 r^s - k_{z1}^2 r^p & 0 & 0 \\ 0 & k_1^2 r^s - k_{z1}^2 r^p & 0 \\ 0 & 0 & 2k_\rho^2 r^p \end{bmatrix}\begin{bmatrix} p_x \\ p_y \\ p_z \end{bmatrix} e^{ik_{z1}(z+h)} dk_\rho. \quad (S9)$$

by substituting Eq. (S9) in the expression for the force (S7), we can obtain the exact integral for the time-averaged force on the dipole

$$\langle F_z(t) \rangle = \frac{1}{2}\Re\left\{ \frac{-1}{8\pi\varepsilon_1}\int_0^\infty k_\rho \left[ \left(|p_=|^2\right)\left(k_1^2 r^s - k_{z1}^2 r^p\right) + |p_\perp|^2 \left(2k_\rho^2 r^p\right) \right] e^{ik_{z1}(2h)} dk_\rho \right\}, \quad (S10)$$

in which we have divided the dipole moment into horizontal $|p_=|^2 = |p_x|^2 + |p_y|^2$ and vertical $|p_\perp|^2 = |p_z|^2$ components. This integral for the force depends, among other factors, on the frequency. In an effort to make this expression frequency independent (aside from the term $h/\lambda$), we divided this force by the power radiated by the dipole in the unbounded medium, rendering a general expression for the force per unit radiated power which turns out to be independent of the frequency as long as the height $h$ is kept constant relative to the wavelength $\lambda_1$. To show this, we consider the following normalization of the variable of integration $\eta = k_\rho / k_1$ such that $k_{z1} = k_1\sqrt{1-\eta^2}$:

$$\langle F_z(t) \rangle = \frac{1}{2}\Re\left\{ \frac{-k_1^4}{8\pi\varepsilon_1}\int_0^\infty \eta \left[ \left(|p_=|^2\right)\left(r^s - (1-\eta^2) r^p\right) + |p_\perp|^2 \left(2\eta^2 r^p\right) \right] e^{ik_1\sqrt{1-\eta^2}(2h)} d\eta \right\}. \quad (S11)$$

One caveat to be considered when evaluating the integral numerically is that the singularity now happens at $\eta = 1$ rather than at $k_1$. Further by substitutions $k_1^2 = \omega^2 \mu_1 \varepsilon_1$, $c_1^2 = 1/\mu_1\varepsilon_1$, $k_1 h = 2\pi(h/\lambda_1)$ and considering that the power radiated by a dipole is given by $P_{rad} = |\mathbf{p}|^2 \omega^4 / (12\pi\varepsilon_1 c_1^3)$, we finally arrive at:

$$\langle F_z(t) \rangle = -\frac{3P_{rad}}{4c_1}\Re\left\{ \int_0^\infty \eta \left[ F_=\left(r^s - (1-\eta^2) r^p\right) + F_\perp \left(2\eta^2 r^p\right) \right] e^{i4\pi\left(\frac{h}{\lambda_1}\right)\sqrt{1-\eta^2}} d\eta \right\}, \quad (S12)$$

where $F_= = |p_=|^2 / |\mathbf{p}|^2$ and $F_\perp = 1 - F_= = |p_\perp|^2 / |\mathbf{p}|^2$ are the fractions of the power radiated by the horizontal and vertical components of the electric dipole, respectively. This integral expression is exact for the full dynamic case (no quasistatic nor image theory approximations), it shows that the force is proportional to the radiated power of the dipole in free space, with the constant of proportionality being independent of frequency, and it yields the normalized plots presented in Figs. 2, (b) and (c), valid at any frequency. We also present, in Fig. S1, (a) and (b), the normalized force for a vertical electric dipole (VED), very similar to the horizontal electric dipole (HED).

*Quasistatic approximation and image theory*

In order to deal with simpler expressions, we apply the quasistatic approximation to Eq. (S12). This approximation is best if the dipole is very close to the substrate (compared to the wavelength)—this is the case under study for strong repulsive forces. According to the quasistatic approximation, we assume $k_1 \ll k_\rho$, implying that $k_{z1} \approx k_{z2} \approx ik_\rho$. This assumption can be translated into the Fresnel reflection coefficients $r^p$ and $r^s$ so they are approximated by a constant value (given by their limit when $k_\rho \to \infty$)

$$r^s(k_\rho) \approx r^s_{qs} = \frac{\mu_2 - \mu_1}{\mu_2 + \mu_1}, \quad (S13a)$$

$$r^p(k_\rho) \approx r^p_{qs} = \frac{\varepsilon_2 - \varepsilon_1}{\varepsilon_2 + \varepsilon_1}, \quad (S13b)$$

This means that the reflection coefficients are considered constant and therefore can be taken out of the integral. By substituting (S13) into (S12), one finds that the resulting integral has an analytic solution. The resulting force has many terms involving different powers of the height ($h/\lambda_1$), each of them multiplied by either $Sin(4\pi h/\lambda_1)$ or $Cos(4\pi h/\lambda_1)$. A further simplification can be made when $(h/\lambda_1) \ll 1$, resulting in the expression for low heights presented in the main text [Eq. (1)]. To show the accuracy of this approximation, we plot the results obtained from Eq. (1) for a VED in Figs. S1, (c) and (d). Clearly the results are very similar to the results obtained from the exact solutions, in Figs. S1, (a) and (b).

We can gain more insight into the approximation by applying (S13) to the electric field (S8), rather than directly to the force, to obtain an analytic expression for the field, and then derive the force using (S7). Consider for simplicity a VED ($F_\perp = 1$); the electric field in the z-axis (S8) under the quasistatic approximation has an analytic solution as follows

$$\mathbf{E}^{VED}_{\text{sec}}(\mathbf{r} = z\hat{\mathbf{z}}) \approx \hat{\mathbf{z}} \frac{-p_z r^p_{qs}}{4\pi\varepsilon_1} \int_0^\infty \frac{k_\rho^3}{ik_{z1}} e^{ik_{z1}(z+h)} dk_\rho = \hat{\mathbf{z}} \cdot r^p_{qs} \frac{p_z}{4\pi\varepsilon_1} \left( \frac{2}{(z+h)^3} - \frac{2ik_1}{(z+h)^2} \right) e^{ik_1(z+h)} \quad (S14)$$

This electric field is equal to that of an image dipole located at $z = -h$ with a dipole moment $p'_z = r^p_{qs} p_z$. This is exactly consistent with image theory applied to static charges over a dielectric medium[24], so the quasistatic approximation leads us to a simple image theory. The force on the VED can then be calculated by applying Eq. (S7). A similar procedure can be conducted for the HED, also involving an image dipole.

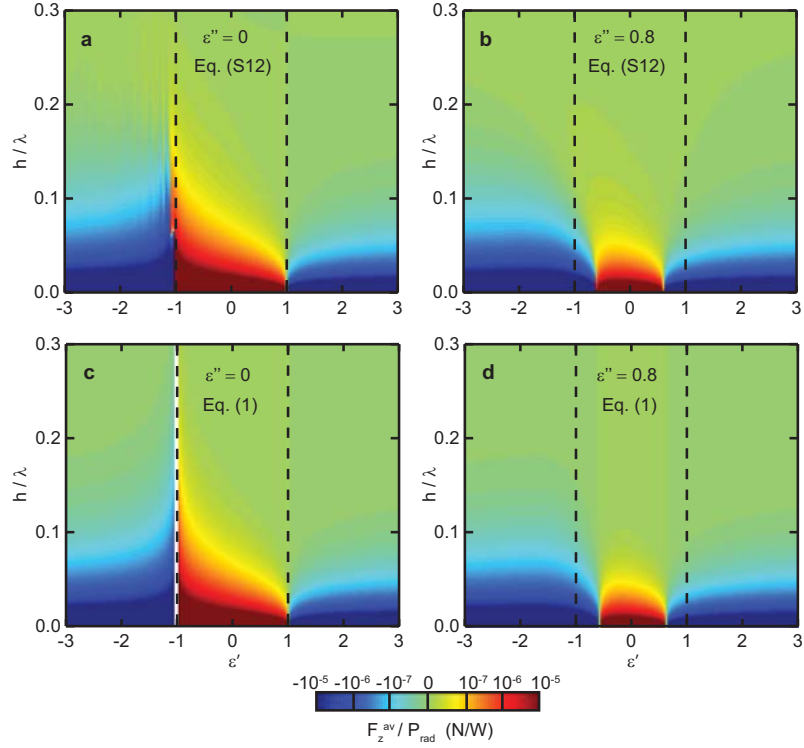

**Figure S1.** *Time-averaged vertical force acting on a vertical electric dipole, normalized to the radiated power of the dipole in the unbounded medium, as a function of height and ε'. Exact electrodynamic calculation for (a) ε'=0 and (b) ε''=0.8. Approximate expression given in Eq. (1) for (a) ε'=0 and (b) ε''=0.8 showing remarkable similarity.*

*Equivalence of an ENZ substrate to a perfect magnetic conductor surface*

For a VED over a lossless ENZ, the quasistatic approximation used in (S14) is no longer an approximate solution, but becomes exactly equal to the full solution. The fields created by a VED can be decomposed into p-polarized plane waves alone (with **H** tangential to the interface), and therefore the force acting on a VED depends only on the Fresnel reflection coefficient of p-polarized waves, $r^p$, whereas the force on a HED depends on both $r^p$ and $r^s$. This can be clearly seen in expression (S12). It turns out that $r^p$, according to its definition from Eq. (S5b), becomes exactly $r^p = r_{qs}^p = -1$ when the substrate is a lossless ENZ ($\varepsilon_2 = 0$), and therefore in that case there is no difference between the quasistatic approximation and the exact solution. The force acting on the VED, which depends only on $r^p$, can therefore be calculated analytically. Furthermore, the solution corresponds to an image dipole of opposite value located at $z = -h$ responsible for the repulsive force. For the fields external to the ENZ substrate, this is exactly equivalent to using a perfect magnetic conductor (PMC) surface instead of an ENZ substrate, which could be an interesting alternative in a microwave regime demonstration. This is also a good method to simplify numerical simulations of a VED over a lossless ENZ substrate. For the HED, the equivalence between an ENZ medium and a PMC surface is not exact due to the presence of s-polarized waves (unless we also make $\mu_2 = 0$, which may not be feasible in practice), but is still a good approximation.

*Numerical simulations of the force on the electric dipole*

To check our analytical results, we have computed the force acting on the dipole using numerical simulations. To do so, we used the software package CST Microwave Studio[25], which uses a finite element method to solve Maxwell's equations for a given structure. The method is a generalization of FDTD. In our simulations of the HED, we used a Lorentzian model $\varepsilon_r(\omega) = \varepsilon_\infty + \omega_0^2(\varepsilon_s - \varepsilon_\infty)/(\omega_0^2 + i\omega\delta - \omega^2)$ ($\varepsilon_\infty$ = 6.7, $\varepsilon_s$ = 10, $\omega_0$ = $2\pi \cdot 23.79 \times 10^{12}$ rad s$^{-1}$, $\delta$ = $2\pi \cdot 0.1428 \times 10^{12}$ rad s$^{-1}$)—parameters analogous to those of SiC[23]—for the permittivity of the substrate, showing $\varepsilon = 0 + i\, 0.1$ at 29.06 THz. We performed all the simulations at that frequency. For the VED, we made use of the fact that a lossless ENZ substrate is exactly equivalent to a PMC surface to greatly reduce the simulation time. The dipole must have a finite length in the simulation, so we used a dipole of length $L$ = 20 nm = $\lambda/516$ and a current $I$ = 1 A. The dipole moment is then given as $\mathbf{p} = I\mathbf{L}/(-i\omega)$. As we will see, a large enough simulation region must be considered and should be finely meshed. Once the electric and magnetic fields are computed for a given height of the dipole above the substrate, we have to estimate the acting force. To do this, we used Maxwell's stress tensor. It is known that the total force acting on any material objects can be found by calculating the integral of Maxwell's stress tensor on any surface that defines a volume containing the objects [19]:

$$\langle \mathbf{F}(t) \rangle = \iint_S \langle \bar{\bar{\mathbf{T}}}(\mathbf{r},t) \rangle \cdot \hat{\mathbf{n}} \cdot dS, \tag{S15}$$

where $\bar{\bar{\mathbf{T}}}(\mathbf{r},t)$ is Maxwell's stress tensor, $S$ is the surface surrounding the volume containing the objects and $\hat{\mathbf{n}}$ is the vector normal to the surface. In our simulations, we are interested in calculating the force acting on the dipole, so we choose the volume to be a 2D slab with its two surfaces perpendicular to the z-axis [See Fig. 2S(a)] so that the volume contains the dipole. One of the surfaces is below the dipole $z = z_{\text{below}} < h$ and the other surface is above the dipole at $z = z_{\text{above}} > h$. This means that the normal vector will be $-\hat{\mathbf{z}}$ and $+\hat{\mathbf{z}}$ on each of the surfaces, respectively. Considering this fact and that due to symmetry considerations the total force is along z-axis, we only need to calculate the $\langle T_{zz}(\mathbf{r},t) \rangle$ element of the time-averaged stress tensor, given as

$$\langle T_{zz}(\mathbf{r},t) \rangle = \frac{1}{2}\varepsilon_1\left(|E_z|^2 - \frac{|E|^2}{2}\right) + \frac{1}{2}\mu_1\left(|H_z|^2 + \frac{|H|^2}{2}\right). \tag{S16}$$

The time-averaged force can then be calculated by applying (S15) to the chosen surfaces, resulting in the subtraction of two integrals of a scalar function over an infinite x-y plane. In practice, since the electromagnetic fields decay away from the dipole, so does $\langle T_{zz}(\mathbf{r},t) \rangle$, so the integration on infinite planes of constant z-value can be computed by considering a finite large enough area in the x-y plane.

$$\langle F_z(t) \rangle = \iint_{z_{\text{above}}} \langle T_{zz}(\mathbf{r},t) \rangle \cdot dS - \iint_{z_{\text{below}}} \langle T_{zz}(\mathbf{r},t) \rangle \cdot dS. \tag{S17}$$

For convenience, we define the following function

$$A(z_0) = -\iint_{z=z_0} \langle T_{zz}(\mathbf{r},t) \rangle \cdot dS \tag{S18}$$

so that $\langle F_z(t) \rangle = A(z_{\text{below}}) - A(z_{\text{above}}) = A_{\text{below}} - A_{\text{above}}$. Theoretically, since any value of $z_{\text{above}} > h$ and $0 < z_{\text{below}} < h$ defines a valid volume surrounding the dipole, $A(z)$ should be a piecewise function equal to a constant value $A_{\text{below}}$ for $0 < z < h$ and $A_{\text{above}}$ for $z > h$. $A(z)$ can be calculated numerically from the simulations to check this fact for consistency. Fig. 2SB shows a plot of $A(z_0)$. We can see that $A(z_0)$ is not exactly a piecewise constant function. We have found two reasons for this: first, if the z plane is too close to the dipole ($z_0 \approx h$) then the $\langle T_{zz}(\mathbf{r},t) \rangle$ function shows high spatial frequency components (i.e., it has high values concentrated locally near the z-axis, as seen in Fig. S2(c)), meaning that a numerical integration with a given mesh size will likely produce large numerical errors the closer the plane is to the dipole. Second, if the plane of integration is too far from the dipole, the finite integration region (due to a finite simulation region) can lead to truncation of the $\langle T_{zz}(\mathbf{r},t) \rangle$ function's tails, resulting in an inaccurate calculation of the integral. In summary, we need to choose $z_{\text{above}}$ and $z_{\text{below}}$ such that the plane of integration is not too close to the dipole as to show numerical errors due to an insufficient mesh size, and not too far from the dipole as for the fields outside the simulated region to be non-negligible. Both requirements are alleviated when we use finer meshes and larger simulation regions, respectively: The numerical error decays faster with the distance to the dipole for finer meshes, while the truncation takes place at larger $z_0$ values when the simulation region is large. These two features lead to lengthy simulations. For the integration at a plane $z_{\text{below}}$ we are restricted to the region $(0,h)$, so the mesh has to be fine enough for the numerical errors to be negligible at least at a distance $h$ from the dipole: this imposes a maximum mesh size for consistent results. For simulation shown in Fig. S2(b), we used a very fine mesh $\lambda/100$ such that the value of $A_{\text{below}}$ can be clearly seen above the numerical error for a sufficient distance from the dipole. We used a relatively large simulation region ($2\lambda \times 2\lambda$) so that $A_{\text{above}}$ reaches a constant value before starting to decay at higher $z_0$ due to truncation of the integrand on the finite simulation region. The final value for the force at the given height of the dipole is then computed as $A_{\text{below}} - A_{\text{above}}$. Repeating the simulation for different heights (taking into account all the previous considerations regarding mesh size and simulation region size in each case), we can obtain the data points in the graphs presented in Figs. 3 and S3.

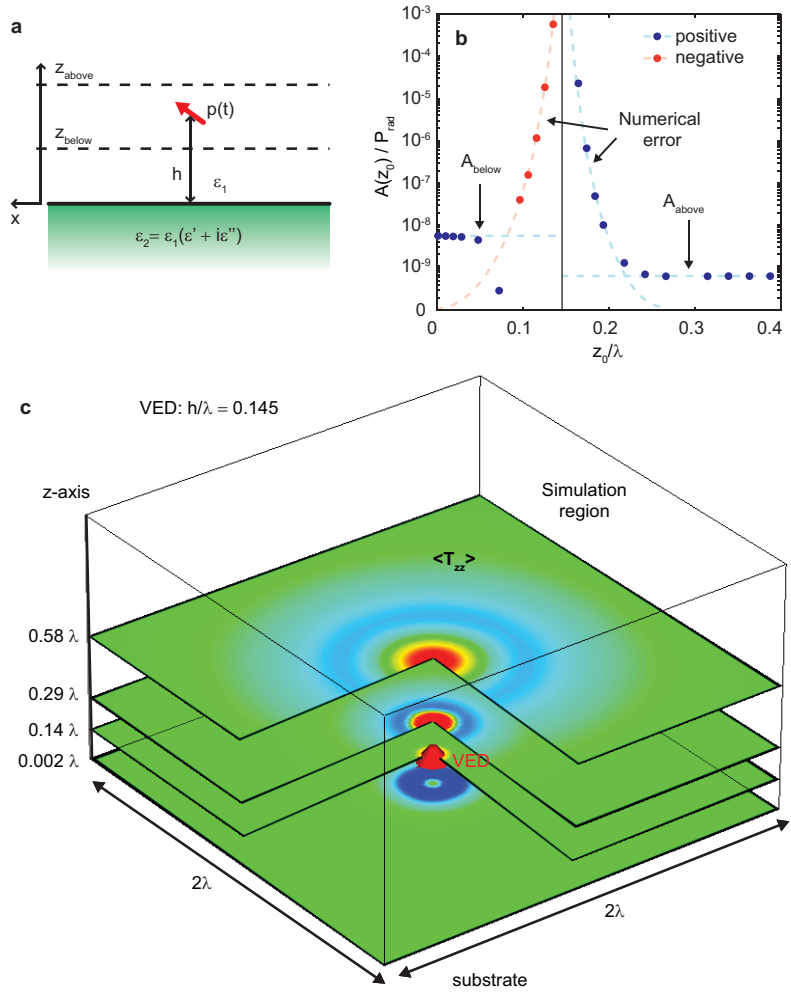

**Figure S2.** *(a) Geometry of the problem showing the two infinite planes of integration at $z = z_{below}$ and $z = z_{above}$. (b) Plot of the numerical integration of the stress tensor at different z planes $A(z_0)/P_{rad}$ for a vertical electric dipole placed at a height $h = 0.145\lambda$ above a lossless ENZ substrate. The dashed lines are a guide for the eye. The values of $A(z_0)$ can be interpreted as the sum of the theoretically expected constant values above and below the dipole with the integration numerical error, which is greater for z-planes closer to the dipole. The numerical error can be seen to decrease (not shown) with finer mesh. (c) Plot of $<T_{zz}>$ at different $z_0$ planes for a vertical dipole placed at a height $h = 0.145\lambda$. The color scale was adjusted independently at each plane for visualization purposes. The plane at $z = 0.14\lambda$ is very close to the dipole and can be seen to have $<T_{zz}>$ with very high spatial frequencies, which gives rise to integration errors.*

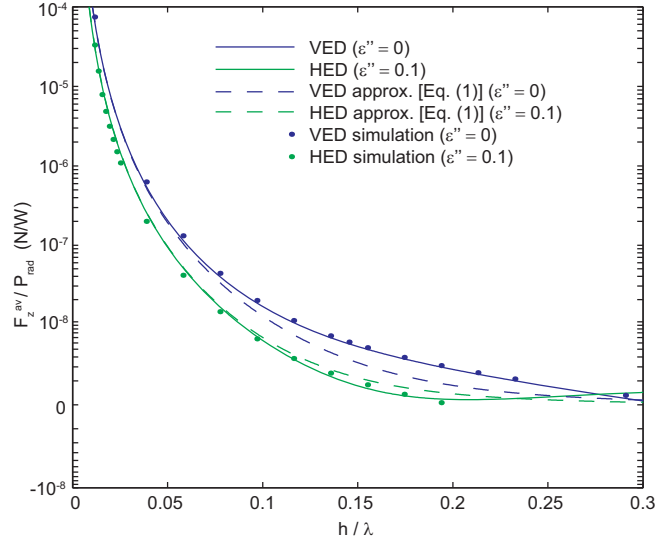

**Figure S3.** *Plot of the time-averaged force per unit radiated power as a function of normalized height above an ENZ substrate. The exact numerical integration [Eq. (S12)], the simple approximation [Eq. (1)], and the numerically simulated values (using CST Microwave Studio to compute Maxwell's stress tensor) are compared for a vertical dipole over a lossless substrate and a horizontal dipole over a lossy ENZ substrate. Excellent agreement is observed between theory and simulations, and agreement with the approximate formula [Eq. (1)] can be seen for small values of h/λ.*

*Calculations of the levitation condition for practical examples:*

The levitation condition for this problem can be stated as:

$$F_z^{\text{avg}} > mg = \rho V g, \tag{S19}$$

where $F_z^{\text{avg}}$ is the time-averaged vertical levitating force given by Eq. (S12) [and approximated by Eq. (1)], $mg$ is the weight of the dipole, $m$ is mass, $\rho$ is mass density, $V$ is volume, and $g = 9.81\ ms^{-2}$ is the acceleration due to gravity. Using the facts that $|\mathbf{p}| = |I|L/\omega$ and $P_{rad} = |\mathbf{p}|^2 \omega^4/(12\pi\varepsilon_0 c^3)$, which we can substitute into Eq. (1), we can calculate $F_z^{\text{avg}}$ and solve Eq. (S19) for the electric current $I$. The required condition on the current for levitation can be derived as following

$$I > k \frac{\sqrt{m}(h/\lambda_1)^2}{(L/\lambda_1)}, \tag{S20}$$

where $k = \sqrt{(512/3)\pi^3 \varepsilon_1 c_1^2 g}$. As an example let us consider a microwave application. Among others, we need to know the mass of the antenna. So as an approximation, let us consider a λ/4 dipole antenna, with radius $R = \lambda/100$ and $L = \lambda/4$, so we can deduct that $m = \rho V = \rho\pi R^2 L$, and since the mass-density of copper $\rho$ is 8.92 g/cm³, we can obtain the expression presented in the report as the minimum required current for levitation at a given height.

The same inequality (S19) is used for the calculation of the fluorescent particle, but this time Eq. (1) is used directly for the calculation of $F_z^{ave}$ as a function of $P_{rad}$, and knowing that the volume of the spherical particle is $V = 4\pi R^3/3$, we can obtain the minimum pump power required for levitation.

*Self-consistent excitation and levitating force on a polarizable particle:*

The calculations so far have been performed for a dipole source of fixed dipole moment. If a polarizable particle is used instead as the levitating dipole, it can be polarized by an external incoming wave, which will also exert a force on the particle. Therefore, the total force, derived from Eq. (S6b), can be split into two terms, the first term is the force caused by the secondary fields of the dipole (reflected on the substrate) on itself, which has been the main result discussed in this paper [Eq. (S10) which is approximated by Eq. (1)], while the second term is the force due to the plane wave. In the case of a horizontal dipole excited by a normally incident plane wave, the two terms are given by:

$$\langle F_z \rangle = \|p_x\|^2 \langle F_{z,px=1}^{dipole} \rangle + \frac{1}{2} \text{Re}\left\{ p_x^* \cdot \frac{\partial E_{pw,x}}{\partial h} \right\} \tag{S21}$$

where $h$ is the height above the substrate, $p_x$ is the dipole moment of the polarizable particle, $E_{pw,x} = \sqrt{2\eta_1 P_{inc}} (e^{-ik_1 h} + r e^{+ik_1 h})$ is the electric field of the incoming and reflected plane wave, $r$ is the reflection coefficient for normal incidence in the substrate, and $\langle F_{z,px=1}^{dipole} \rangle$ can be obtained by substituting $p_{//} = 1$ and $p_\perp = 0$ in Eq. (S10). Notice that the first term (the dipole force) is proportional to the square of the dipole moment magnitude, while the second term (the plane wave force) is linearly proportional to the dipole moment. Thus the dipole force dominates over the plane wave force for strongly polarized particles. To obtain the polarization of the particle we must apply the relation $\boldsymbol{p} = \alpha \boldsymbol{E}$ where α is the polarizability of the particle and $\mathbf{E}$ is the total minus the self-generated electric field at the position of the particle, given by the incoming and reflected plane wave field $E_{pw}$ added to the secondary fields of the dipole $E_{sec}$ (reflected on the substrate). This results in a self-consistent dipole moment satisfying $p_x = \alpha(E_{pw,x} + E_{sec,x}) = \alpha(E_{pw,x} + p_x E_{sec,x,p=1})$, which can be written as:

$$p_x = \frac{\alpha E_{pw,x}}{1 - \alpha E_{sec,x,p=1}}, \tag{S22}$$

where $E_{sec,x,p=1}$ is the reflected dipole field created by a dipole with unit dipole moment, oriented along the x direction, obtained by substituting $p_x = 1$ and $p_y = p_z = 0$ into Eq. (S8). The main effect of this self-consistent excitation is that, when the dipole is very close to the surface, its own reflected fields tend to reduce its polarization, so the polarization of the dipole tends to zero as the dipole gets closer to the surface, i.e. $p_x(h \to 0) \to 0$. This ensures that the first term in the force Eq. (S21) remains finite, despite the factor $\langle F_{z,px=1}^{dipole} \rangle$ going to infinity close to the surface, as discussed in the paper. We should clarify that the possibility of a resonant polarization, predicted when the denominator of Eq. (S22) goes to zero, has not appeared in any of the cases we have studied, because $E_{sec,x,p=1}$ is negative at very low heights, while at bigger heights it is oscillating around zero, but its magnitude is generally smaller than $1/\alpha$. The polarizability of a non-resonant electrically small particle in free space is given approximately by:

$$\alpha \approx \frac{1}{3} V \varepsilon_0 \frac{\varepsilon_p - 1}{\varepsilon_p + 2}, \tag{S23a}$$

where $\varepsilon_p$ is the particle permittivity and $V$ is the particle volume, while the polarizability of a resonant particle is related to the scattering coefficient $C_1$ [S1] through:

$$\alpha \approx \frac{-6\pi i \varepsilon_0}{k_0^3} C_1, \tag{S23b}$$

where $C_1$ goes up to 1 at resonance. Combining Eqs (S21 – S23) we have all the ingredients necessary to calculate the total force acting on a polarizable particle when a plane wave is incident normal to the substrate. The general case of a plane wave incident at any angle could be calculated similarly but requires vector versions of Eqs. (S21) and (S22).

The high number of parameters that exist in this problem gives rise to a big range of possible options. A thorough study of all the possibilities is out of the scope of this paper, but here we show a particular instance of a levitating polarizable particle. Consider a silver nanosphere of radius 20 nm (density of silver $\rho$ = 10490 kg/m$^3$). The real part of its permittivity achieves the resonance condition $\text{Re}(\varepsilon_{Ag}) = -2$ around 845 THz [S2]. Taking the polarizability at that frequency equal to Eq. (S23b) with $C_1$ = 1, considering an ENZ metamaterial with $\varepsilon = 0 + 0.5i$ at 845 THz being used as a substrate, and considering a normal incident plane wave coming from a 5mW laser focused on an area of 1mm$^2$ (comparable with the power density of a common laser pointer), the achieved ratio of levitating force to weight is shown in Fig. S4, computed numerically using Eqs. (S8), (S10), (S21) and (S22). In this example, a ratio greater than unity (required for levitation) is achieved at heights between 0.07$\lambda$ and 0.11$\lambda$, which means that the particle would show stable levitation at a height of 39 nm, almost equal to its diameter.

*Supplementary References:*
[S1] Alù, A. & Engheta, N. *Journal of Applied Physics* **97**, 094310 (2005).
[S2] Johnson, P. B. & Christy, R. W. *Physical Review B* **6**, 4370–4379 (1972).

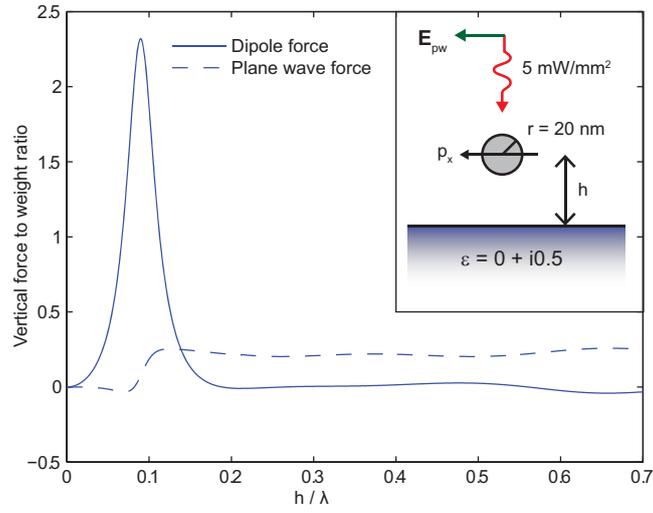

Fig. 4S. *Self-consistent calculation of the force-to-weight ratio of a 20 nm radius silver nanosphere at resonance (845 THz) with polarizability $\alpha \approx -6\pi i \varepsilon_0 / k_0^3$ suspended over an ENZ metamaterial with $\varepsilon = 0 + 0.5i$, under illumination of an incident plane wave with a power density of 5 mW/mm². The inset shows a depiction of the proposed excitation.*